\documentclass[twocolumn,showpacs,floatfix,superscriptaddress,amsmath,amssymb,prl]{revtex4}
\usepackage{amssymb}
\usepackage{graphicx}
\usepackage{hyperref}
\usepackage{ulem}

\newcommand{\be}{\begin{equation}}
\newcommand{\ee}{\end{equation}}

\begin{document}

\title{ Renormalization group flows and quantum phase transitions: fidelity versus entanglement}

\author{Huan-Qiang Zhou}
\affiliation{Centre for Modern Physics and Department of Physics,
Chongqing University, Chongqing 400044, The People's Republic of
China}
\begin{abstract}
We compare the roles of fidelity and entanglement in characterizing
renormalization group flows and quantum phase transitions. It turns
out that the scaling parameter extracted from fidelity for different
ground states succeeds to capture nontrivial information including
stable and unstable fixed points, whereas the von Neumann entropy as
a bipartite entanglement measure (or equivalently, majorization
relations satisfied by the spectra of the reduced density matrix
along renormalization group flows) often fails, as far as the
intrinsic irreversibility-information loss along renormalization
group flows-is concerned. We also clarify an intimate connection
between the von Neumman entropy, majorization relations, and
fidelity. The relevance to Zamolodchikov's c theorem is indicated.
\end{abstract}
\pacs{03.67.-a,05.70.Fh, 64.60.Ak}

\date{\today}
\maketitle

Quantum phase transitions (QPTs)~\cite{sachdev} are a topic of
current interest, subject to intense study in condensed matter
systems. Conventionally, a QPT is described in terms of a
Hamiltonian and its spectrum. Most examples of the well-studied QPTs
fit into the Landau-Ginzburg-Wilson paradigm, with the central
concept being a local order parameter. The non-zero expectation
value of a local order parameter characterizes a symmetry-breaking
phase that only occurs for a system with infinite number of degrees
of freedom, in contrast to QPTs resulting from a level crossing in a
finite size system.
However, there exist phases beyond
symmetry-breaking orders, which yields exotic QPTs characterized by
topological/quantum orders~\cite{wen}. Unconventional QPTs also
occur in matrix product systems~\cite{wolf}.

Another picture emerges due to latest advances in quantum
information science, which attempts to characterize QPTs from both
entanglement~\cite{preskill,osborne,vidal,entanglement,levin} (for a
review, see~\cite{amico}) and fidelity~\cite{zanardi,zjp,more}
perspectives. Remarkably, for quantum spin chains, the von Neumann
entropy, as a bipartite entanglement measure, exhibits qualitatively
different behaviors at and off criticality~\cite{vidal}. On the
other hand, it has been shown that fidelity may be used to
characterize QPTs, regardless of what type of internal order is
present in quantum many-body states~\cite{zjp}. A basic point is
that there always exists an order parameter (local/nonlocal) in a
system undergoing a QPT. In principle, such an order parameter may
be constructed systematically~\cite{oshikawa}. This results in the
orthogonality of different ground states, due to state
distinguishability~\cite{nielsen}. Thus, it is justified that the
scaling parameter may be extracted from fidelity to characterize
QPTs. In fact, nontrivial information about QPTs including stable
and unstable fixed points along renormalization group (RG) flows may
be revealed solely from ground states, and critical exponents can be
extracted by performing a finite size scaling analysis from such a
scaling parameter~\cite{zzl}.

An intriguing question is to clarify the connection  between
fidelity and entanglement approaches to QPTs. An apparent difference
between entanglement measures and fidelity is that entanglement
measures involve {\it partitions}, whereas the fidelity approach
treats systems as a {\it whole}.  Therefore, {\it no matter what
bipartite entanglement measures are used, some physical information
will be lost,  since the whole is not simply the sum of the parts}.
On the other hand, one may expect that the scaling parameter
extracted from fidelity should capture information that is lost in
bipartite entanglement measures.

In this paper, we compare the roles of fidelity and entanglement in
characterizing RG flows and QPTs. It is found that the scaling
parameter extracted from fidelity for different ground states
succeeds to capture nontrivial information including stable and
unstable fixed points, whereas the von Neumann entropy as a
bipartite entanglement measure (or equivalently, majorization
relations satisfied by the spectra of the reduced density matrix
along RG flows) often fails, as far as the intrinsic
irreversibility, i.e., information loss along RG flows is concerned.
We also clarify an intimate connection between the von Neumman
entropy, majorization relations and fidelity. The relevance to
Zamolodchikov's c theorem is indicated, if a system is conformally
invariant at transition points,.

{\it Generalities.} For a quantum spin chain described by a
Hamiltonian $H(\lambda)$, with $\lambda$ a control parameter, it is
well-established~\cite{verstraete} that its ground state
$|\psi(\lambda)\rangle$ may be represented in terms of the so-called
matrix product states (MPS)~\cite{mps}. Suppose the system is
translationally invariant, then $|\psi(\lambda)\rangle$ takes the
form,
\begin{equation}
 |\psi(\lambda)\rangle = {\rm Tr} (A_{i_1}(\lambda) \cdots A_{i_L}(\lambda)) |i_1 \cdots i_L \rangle, \label{MPS}
\end{equation}
where $ \{ A_i \}$ is a set of $d D \times D$ matrices, with $d$
being the dimension of the local Hilbert space at each lattice site,
and $D$ the dimension of the bonds in the valence bond
picture~\cite{verstraete,wolf}. The fidelity $F(\lambda, \lambda')
\equiv |\langle \psi (\lambda') |\psi(\lambda)\rangle|$ for two
ground states $|\psi (\lambda)\rangle$ and $|\psi(\lambda')\rangle$
corresponding to different values of the control parameter $\lambda$
is
\begin{equation}
F(\lambda, \lambda')  = {\rm Tr} (E^L(\lambda,\lambda')) ,
\label{MPS}
\end{equation}
with the (generalized) transfer matrix $E(\lambda,\lambda') = \sum
_{s=1}^d A^{s*}(\lambda') \otimes A^s(\lambda)$. Therefore, it is
straightforward to evaluate the fidelity $F$ by diagonalizing the
transfer matrix $E(\lambda,\lambda')$. For large enough $L$'s, the
fidelity $F$ scales as $F(\lambda,\lambda') \sim
d^L(\lambda,\lambda')$, where $d(\lambda,\lambda')$ is the scaling
parameter. In the thermodynamic limit, $d(\lambda,\lambda')$ may be
defined as
\begin{equation}
\ln d(\lambda,\lambda') = \lim_{L \rightarrow \infty} \ln
F(\lambda,\lambda') /L.
\end{equation}
In fact, the scaling parameter  $d(\lambda,\lambda)$ is the largest
eigenvalue of the transfer matrix $E(\lambda,\lambda')$, if there is
no pair of eigenvalues that are complex conjugate each other
(otherwise, a fast oscillating part should first be factored
out~\cite{zjp}). Obviously, it satisfies the properties inherited
from the fidelity $F(\lambda,\lambda')$: (a) $d(\lambda,\lambda)=1$;
(b) $d(\lambda,\lambda')=d(\lambda',\lambda)$; and (c) $0 \leq
d(\lambda,\lambda')\leq1$.

Although the system is in a pure state, a subsystem $A$ consisting
of $N$ consecutive spins is generically in a mixed state described
by the reduced density matrix $\rho _A(\lambda) = {\rm Tr}_B \rho
(\lambda)$, with the density matrix $\rho (\lambda) = |\psi
(\lambda) \rangle \langle \psi (\lambda ) |$, and the trace is taken
over the subsystem $B$ complementary to the subsystem $A$. The
fidelity $F_A(\lambda,\lambda')$ for two reduced density matrices
$\rho_A(\lambda)$ and $\rho_A(\lambda')$ is defined as $
F_A(\lambda, \lambda') ={\rm Tr} \sqrt{\rho^{1/2}_A (\lambda) \rho_A
(\lambda')\rho^{1/2}_A (\lambda)}$.  For large enough $L$'s and
$N$'s, with a fixed $N/L$~\cite{ratio}, $F_A$ scales as
$F_A(\lambda,\lambda') \sim d_A^{N}(\lambda,\lambda')$, where
$d_A(\lambda,\lambda')$ is the scaling parameter for the subsystem
$A$. Formally, in the thermodynamic limit, $d_A(\lambda,\lambda')$
may be defined as
\begin{equation}
\ln d_A(\lambda,\lambda') = \lim_{N \rightarrow \infty} \ln
F_A(\lambda,\lambda') /N.
\end{equation}
The scaling parameter  $d_A(\lambda,\lambda)$ also enjoys the
properties: (a) $d_A(\lambda,\lambda)=1$; (b)
$d_A(\lambda,\lambda')=d_A(\lambda',\lambda)$;  and (c) $0 \leq
d_A(\lambda,\lambda')\leq1$. The scaling parameters
$d(\lambda,\lambda')$ and $d_A(\lambda,\lambda')$ satisfy
\begin{equation}
d_A(\lambda,\lambda') \ge d(\lambda,\lambda'). \label{neq}
\end{equation}
This follows from the well-known inequality: $F_A(\lambda,\lambda')
\ge F(\lambda,\lambda')$~\cite{nielsen}.  Although it is
straightforward to calculate $d(\lambda,\lambda')$ in the MPS
representation, it is a formidable task to evaluate
$d_A(\lambda,\lambda')$.  This is due to the fact that
$d_A(\lambda,\lambda')$ depends not only on the spectra but also on
the eigenvectors of the reduced density matrices $\rho_A(\lambda)$
and $\rho_A(\lambda')$. From the Schmidt decomposition, one may
expect that $d_A(\lambda,\lambda')$ takes the form
$d_A(\lambda,\lambda')=\sum _{\alpha,\beta} g_{\alpha
\beta}(\lambda,\lambda')\sqrt{\omega_\alpha(\lambda)}\sqrt{\omega_\beta(\lambda')}$,
with $\alpha, \beta =1,\cdots, D$ and $\{ w_\alpha (\lambda) \}$ the
(normalized and non-increasing ordered) spectra of the reduced
density matrix $\rho_A$, and $g_{\alpha \beta}(\lambda,\lambda')$ a
model-dependent real tensor, satisfying $g_{\alpha
\beta}(\lambda,\lambda')=g_{\beta, \alpha}(\lambda',\lambda)$. The
difficulty arises from the dependence of $g_{\alpha
\beta}(\lambda,\lambda')$ on $\lambda$ and $\lambda'$. Thus, instead
of $d_A(\lambda,\lambda')$, we may seek a quantity
$d_E(\lambda,\lambda')=\sum _{\alpha,\beta} E_{\alpha \beta}
\sqrt{\omega_\alpha(\lambda)}\sqrt{\omega_\beta(\lambda')}$, which
is universal in the sense that the real tensor $E_{\alpha \beta}$ is
independent of $\lambda$ and $\lambda'$.  The only choice is
$E_{\alpha \beta}=\delta_{\alpha \beta}$, if one requires that
$d_E(\lambda,\lambda')$ enjoys the same properties as
$d(\lambda,\lambda')$ and $d_A(\lambda,\lambda')$: (a)
$d_E(\lambda,\lambda)=1$; (b)
$d_E(\lambda,\lambda')=d_E(\lambda',\lambda)$;  and (c) $0 \leq
d_E(\lambda,\lambda')\leq1$. Thus $d_E(\lambda,\lambda')=
\sum_{\alpha}\sqrt{\omega_\alpha(\lambda)}\sqrt{\omega_\alpha(\lambda')}$.
We stress that $d_E(\lambda,\lambda')$ only involves the eigenvalues
of the reduced density matrix, whereas $d(\lambda,\lambda')$ unveils
information encoded in the entire ground state.

{\it The connection between $d(\lambda,\lambda')$,
$d_E(\lambda,\lambda')$ and majorization.} Suppose a system flows to
two different stable fixed points $\lambda_-$ and $\lambda_+$ under
RG transformations~\cite{cardy}, with an unstable fixed point
$\lambda_c$ as a transition point. An intriguing fact is that
$d_E(\lambda,\lambda')$ shares similar behaviors to
$d(\lambda,\lambda')$,  if $\lambda$ and $\lambda'$ are in the same
phase. To demonstrate this, we notice that, generically, there is
one and only one extreme point for $d_E(\lambda,\lambda')$ if one
regards  $d_E(\lambda,\lambda')$ as a function of $\lambda$ for a
given $\lambda'$. That is, when $\lambda=\lambda'$, it reaches the
maximum $1$, as follows from the Lagrange multipliers for
$d_E(\lambda,\lambda')$ under the constraints: $\sum_\alpha
\omega_\alpha (\lambda) =1$ and $\sum_\alpha \omega_\alpha
(\lambda')=1$. Therefore, $d_E(\lambda,\lambda')$ increases with
$\lambda$, until it reaches $1$ when $\lambda =\lambda'$, then it
decreases with $\lambda$ along an RG flow for a fixed $\lambda'$.

We stress that the monotonic behaviors of $d_E(\lambda,\lambda')$
are consistent with majorization relations for the spectra of the
reduced density matrix along RG flows. It has been established that
there is a more ``fine-grained" characterization of entanglement
loss along RG flows in terms of
majorization~\cite{latorre,orus,zbfs}, which states that $m_k \equiv
\sum ^k_{\alpha=1} \omega_{\alpha} (k=1,2,\cdots)$ is non-decreasing
along RG flows. As shown in Ref.~\cite{zbfs}, this simply follows
from the fact that there is one and only one crossing point
$\alpha^*$ for eigenvalue distributions when one regards
$\omega_{\alpha}$ as a function of the index
$\alpha$~\cite{explanation}. Equivalently, $\triangledown_\lambda
\omega_\alpha \geq 0$ for $\alpha \leq \alpha^*$, and
$\triangledown_\lambda \omega_\alpha \leq 0$ for $\alpha >
\alpha^*$, where $\triangledown_\lambda$ denotes the derivative with
respect to $\lambda$ along RG flows. Then, suppose $\lambda$ and
$\lambda'$ are in the same phase and $\lambda'$ is fixed,
$\triangledown_\lambda d_E(\lambda,\lambda') =1/2( \sum_{\alpha \leq
\alpha^*} \sqrt {\omega_\alpha (\lambda')/\omega_\alpha (\lambda)}
\triangledown_\lambda \omega_\alpha (\lambda)+ \sum_{\alpha >
\alpha^*} \sqrt {\omega_\alpha (\lambda')/\omega_\alpha (\lambda)}
\triangledown_\lambda
\omega_\alpha(\lambda))~\geq~1/2\triangledown_\lambda (\sum_\alpha
\omega_\alpha(\lambda))=0$, if $\lambda$ flows from $\lambda_c$ to
$\lambda'$, because $\omega_\alpha (\lambda')/\omega_\alpha
(\lambda) \geq 1$ for $\alpha \leq \alpha^*$, and  $\omega_\alpha
(\lambda')/\omega_\alpha (\lambda) \leq 1$ for $\alpha  > \alpha^*$;
$\triangledown_\lambda d_E(\lambda,\lambda') =1/2( \sum_{\alpha \leq
\alpha^*} \sqrt {\omega_\alpha (\lambda')/\omega_\alpha (\lambda)}
\triangledown_\lambda \omega_\alpha(\lambda)+ \sum_{\alpha >
\alpha^*} \sqrt {\omega_\alpha (\lambda')/\omega_\alpha (\lambda)}
\triangledown_\lambda \omega_\alpha(\lambda)) \leq 0$, if $\lambda$
flows from $\lambda'$ to a stable fixed point, because
$\omega_\alpha (\lambda')/\omega_\alpha (\lambda) \leq 1$ for
$\alpha \leq \alpha^*$, and  $\omega_\alpha (\lambda')/\omega_\alpha
(\lambda) \geq 1$ for $\alpha  > \alpha^*$.

If $\lambda$ and $\lambda'$ are in different phases, then the
monotonic behaviors of $d_E(\lambda,\lambda')$ with $\lambda$ along
an RG flow for a fixed $\lambda'$ depends on what type of orders
present in systems. For our purpose, we discuss the two extreme
cases: $D=2$ for MPS systems, and $D=\infty$ for spontaneous
symmetry-breaking orders. Since the monotonicity is universal in the
sense that it should remain the same for $\lambda'$'s from the same
phase, therefore we just need to focus on the stable fixed points
$\lambda'=\lambda_{\pm}$. For $D=2$, there are only two different
choices for the reduced density matrix spectra at the stable fixed
points: $\{1,0 \}$ and $\{ 1/2,1/2 \}$, corresponding to unentangled
states and maximally entangled states, respectively. For the former,
$\triangledown_\lambda d_E(\lambda,\lambda') = \triangledown_\lambda
\sqrt{\omega_1(\lambda)} \geq 0$, since $\omega_1(\lambda)$ is
non-decreasing along an RG flow. For the latter,
$\triangledown_\lambda d_E(\lambda,\lambda')= 1/\sqrt{2}
\triangledown_\lambda
(\sqrt{\omega_1(\lambda)}+\sqrt{\omega_2(\lambda)}) =1/(2\sqrt{2})
(1/\sqrt{\omega_1(\lambda)}-1/\sqrt{\omega_2(\lambda)})
\triangledown_\lambda \omega_1 \leq 0$. For $D=\infty$ as occurs for
spontaneous symmetry-breaking orders, a remarkable feature is that
the reduced density matrix spectra at critical points tend to
vanish, leading to the divergence of the von Neumann entropy for a
half-infinite chain. Thus, $D_E(\lambda,\lambda')$ tends to 0 when
$\lambda$ or $\lambda'$ tends to $\lambda_c$. On the other hand,
degeneracies occur in the largest eigenvalue of the reduced density
matrix in symmetry-broken phases, and no degeneracy in symmetric
phase. Therefore, the reduced density matrix spectra are $\{
1/n,\cdots,1/n,0,\cdots \}$ ($n$-degeneracies) for stable fixed
points in symmetry-broken phases and $\{ 1,0,\cdots \}$ for stable
fixed points in symmetric phases. Then, $d_E(\lambda,\lambda')$ as a
function of $\lambda$ for a fixed $\lambda'$ monotonically
increases, if $\lambda'$ is in symmetric phases, whereas
$d_E(\lambda,\lambda')$ increases, until it reaches a maximum, then
decreases, if $\lambda'$ is in symmetry-broken phases.

The fact that $d_E(\lambda,\lambda')$ shares similar monotonic
behaviors to $d(\lambda,\lambda')$ establishes a connection between
entanglement and fidelity approaches to QPTs, since information
unveiled in $d_E$ results from the reduced density matrix spectra
that in turn determine the von Neumann entropy $E(\lambda)=
-\sum_\alpha \omega_\alpha (\lambda)\ln \omega_\alpha (\lambda)$.
The latter is non-increasing along RG flows. If a system is
conformally invariant at transition points, then one may formulate
an entropic version of Zamolodchikov's c theorem~\cite{huerta}.
Therefore, the monotonicity of $d(\lambda,\lambda')$ and
$d_E(\lambda,\lambda')$ is reminiscent of Zamolodchikov's c theorem.

{\it Quantum $XY$ spin 1/2 chain.} The quantum $XY$ spin chain is
described by the Hamiltonian
\begin{equation}
H= -\sum_{j=-M}^M ( \frac {1+\gamma}{2} \sigma^x_j \sigma^x_{j+1} +
\frac {1-\gamma}{2} \sigma^y_j \sigma^y_{j+1}  + \lambda \sigma^z_j
). \label{HXY}
\end{equation}
Here $\sigma_j^x, \sigma_j^y$, and $\sigma_j^z$  are the Pauli
matrices at the $j$-th lattice site. The parameter $\gamma$ denotes
an anisotropy in the nearest-neighbor spin-spin interaction, whereas
$\lambda$ is an external magnetic field. The Hamiltonian (\ref{HXY})
may be diagonalized exactly~\cite{lieb}. In the thermodynamic limit,
$d(\lambda,\lambda';\gamma)$ takes the form:
\begin{equation}
d(\lambda,\lambda';\gamma) = \exp [\frac {1}{2\pi} \int ^\pi_0
d\alpha \ln {\cal F} (\lambda,\lambda';\gamma;\alpha)],
\end{equation}
where ${\cal F} (\lambda,\lambda';\gamma;\alpha) = \cos [\vartheta
(\lambda;\gamma;\alpha)-\vartheta(\lambda';\gamma;\alpha)]/2$, with
$\cos \vartheta (\lambda;\gamma;\alpha) = (\cos \alpha -
\lambda)/\sqrt {(\cos \alpha -\lambda)^2+\gamma^2 \sin^2 \alpha}$.
It was shown~\cite{zjp} that $d(\lambda,\lambda';\gamma)$ exhibits a
pinch point at ($1,1$), i.e., an intersection of two singular lines
$\lambda =1$ and $\lambda'=1$.  For quantum Ising model in a
transverse field ($\gamma=1$), all states for $\lambda >1$ flow to
the product state with all spins aligning in the $z$ direction
($\lambda=\infty$), and all states for $\lambda<1$ flow to the cat
state $(|\leftarrow \cdots \leftarrow \rangle+|\rightarrow
\cdots\rightarrow \rangle)/\sqrt {2}$ ($\lambda=0$). We stress that
$d(\lambda,\lambda';\gamma)$ detects the duality between two phases:
$\lambda >1$ and $\lambda<1$, because $d(\lambda,\lambda';1) =
d(1/\lambda,1/\lambda';1)$~\cite{zzl}.

The exact spectra of the reduced density matrix for a half-infinite
chain take the form~\cite{spectra}:
\begin{equation}
\omega_{\{n_0,n_1,\cdots,n_{\infty}\}}(\lambda;\gamma) =\frac
{e^{-\sum _{j=0}^{\infty} \epsilon _j n_j}}{Z(\lambda;\gamma)},
\end{equation}
where $\epsilon_j = (2j+1) \epsilon$ if $\lambda >1$, and
$\epsilon_j = 2j \epsilon$ if $\lambda <1$, and
$Z(\lambda;\gamma)=(16q/(k^2 k'^2))^{1/24}$ if $\lambda >1$, and
$Z(\lambda;\gamma)=2(16 k^2/(q k'))^{1/12}$ if $\lambda <1$, with $q
\equiv \exp (-\epsilon)$. Here $\epsilon = \pi K(k')/K(k)$, with
$K(k)$ being the complete elliptic integral of the first kind, $k$
the modulus and $k'=\sqrt{1-k^2}$. The relation between the modulus
$k$ and the anisotropy $\gamma$ and the transverse field strength
$\lambda$ is $k=\gamma/\sqrt{\lambda^2+\gamma^2-1}$ if $\lambda >1$
and $k=\sqrt{\lambda^2+\gamma^2-1}/\gamma$ if $\lambda <1$.
Therefore, if $\lambda$ and $\lambda'$ are in the same phase, we may
derive a closed expression for $d_E(\lambda,\lambda';\gamma)$:
\begin{equation}
d_E(\lambda,\lambda';\gamma) = \frac
{Z(\lambda,\lambda';\gamma)}{\sqrt{Z(\lambda;\gamma)
Z(\lambda';\gamma)}},
\end{equation}
where $Z(\lambda,\lambda';\gamma)$ takes the same form as
$Z(\lambda;\gamma)$, i.e., $Z(\lambda,\lambda';\gamma)=(16{\tilde
q}/({\tilde k}^2 {\tilde k}'^2))^{1/24}$ if $\lambda, \lambda'>1$,
and $Z(\lambda,\lambda';\gamma)=2(16 {\tilde k}^2/({\tilde q}
{\tilde k}'))^{1/12}$ if $\lambda, \lambda' <1$, with $ {\tilde
q}=\sqrt {q(\lambda) q(\lambda')}$, and $ {\tilde k}$ determined
from ${\tilde q} = \exp (-{\tilde \epsilon})$, with ${\tilde
\epsilon} = \pi K({\tilde k}')/K({\tilde k})$. However, if $\lambda$
and $\lambda'$ are in different phases,
$d_E(\lambda,\lambda';\gamma)$ is only available in terms of an
infinite sum.

The monotonic behaviors of $d_E(\lambda,\lambda';\gamma)$ as an
example for $D=\infty$ with $Z_2$ broken symmetry are numerically
confirmed. If $\lambda$ and $\lambda'$ are in the same phase,
$d_E(\lambda,\lambda';\gamma)$ takes values very close to
$d(\lambda,\lambda';\gamma)$ when $\lambda$ and $\lambda'$ are away
from the critical point. However, in contrast to
$d(\lambda,\lambda';\gamma)$, $d_E(\lambda,\lambda';\gamma)$ fails
to detect the duality between two phases for quantum Ising model in
a transverse field ($\gamma=1$). This is due to the fact that the
dual unitary transformation connecting the two phases involves
nonlocal operations, which do not affect the fidelity (and so
$d(\lambda,\lambda';\gamma)$), but do change the spectra of the
reduced density matrix (and so $d_E(\lambda,\lambda';\gamma)$). We
note that the von Neumann entropy $E(\lambda)$ is decreasing along
the RG flows,  so it succeeds to detect the stable fixed points:
$\lambda =0$ and $\lambda =\infty$, but it fails to detect the
duality.

{\it Quantum spin 1/2 chain with three-body interactions.}  The
system is described by a Hamiltonian with three-body
interactions~\cite{dual}
\begin{equation}
H= \sum_i 2(g^2 -1) \sigma^z_i \sigma^z_{i+1} - (1+g)^2 \sigma^x_i +
(g-1)^2 \sigma^z_i \sigma^x_{i+1} \sigma^z_{i+2}.
\end{equation}
The peculiarity of the model is that it is not conformally invariant
at the transition point $g_c=0$: in the thermodynamic limit, the
correlation length diverges, and energy gap vanishes, but the ground
state energy remains smooth~\cite{wolf}. As emphasized in
Ref.~\cite{zjp}, the parameter space should be compactified by
identifying $g=+\infty$ with $g=-\infty$, due to the fact that $H(+
\infty) =H(-\infty)$. Since ground states are an MPS with
$A_1=(I-\sigma^z)/2+\sigma^-$ and $A_2=(I+\sigma^z)/2+g
\sigma^+$~\cite{wolf}, one may extract the scaling parameter $d$ as
$d(g,g') = \sqrt {1+|g g'|}/ \sqrt {(1+|g|)(1+|g'|)}$ if $g$ and
$g'$ are in different phases, and $d(g,g') = (1+\sqrt {|g g'|})/
\sqrt {(1+|g|)(1+|g'|)}$ if $g$ and $g'$ are in the same
phase~\cite{zjp}. From this we read off that there are two
transition points: $g=0$ and $\infty$. All states for positive $g$
flow to the product state ($g=1$) with all spins aligning in the $x$
direction, and all states for negative $g$ flow to the cluster
state~\cite{raussendorf} ($g=-1$).

The reduced density matrix spectra for a half-infinite chain are
identical to those of $\rho$, which may be calculate exactly by
exploiting the freedom in the set $\{ A_i \}$ in order to fix the
gauge: $\sum_i A_i A_i^\dagger = I$ and $\sum_i A_i^{\dagger} \rho
A_i = \rho$~\cite{wolf}. For $g>0$, the eigenvalues of the reduced
density matrix are $1/2 \pm \sqrt {g}/(1+g)$, whereas for $g<0$,
both of them are $1/2$. Obviously, for positive $g$, the
majorization relations are satisfied, since the larger eigenvalue
$1/2 + \sqrt {g}/(1+g)$ is monotonically increasing when $g$ varies
from 0 to 1, or varies from $\infty$ to 1, consistent with the fact
that both $g=0$ and $g=\infty$ are unstable fixed points, and $g=1$
is a stable fixed point. However, the trivial spectra for negative
$g$ fail to identify another stable fixed point $g=-1$, in contrast
to $d(g,g')$. One may check that $d_E(g,g')=d(g,g')$ if $g,g'>0$,
and $d_E(g,g')=1$, i.e., $d_E(g.g')$ serves as a trivial upper
bound, if $g,g'<0$. On the other hand, $d_E(g,g') = 1/\sqrt{1+g}$ if
$0<g<1$ and $g'<0$, and $d_E(g,g') = \sqrt{g/(1+g)}$ if $1<g<\infty$
and $g'<0$. In this case, $d_E(g,g')$ is decreasing with $g$ along
an RG flow for a fixed $g'$, consistent with the majorization
relations. The von Neumann entropy $E(g<0)=\ln 2$ also fails to
identify the stable fixed point $g=-1$.

{\it Quantum spin 1 chain with two-body interactions.} The model is
a deformation of the celebrated AKLT model~\cite{mps}, with the
Hamiltonian
\begin{equation}
\begin{split}
H &= \sum_i (2+g^2) {\bf S}_i {\bf S}_{i+1} + 2 ({\bf S_i}{\bf
S}_{i+1})^2+ 2 (4-g^2) (S^z)^2\\ &+ (g+2)^2 (S^z_i
S^z_{i+1})^2+g(g+2) \{S^z_i S^z_{i+1}, {\bf S}_i {\bf S}_{i+1} \}_+.
\end{split}
\end{equation}
Note that the AKLT model corresponds to $g=-2$. The model is very
similar to the previous one, although the order parameter is
non-local. Indeed, the parameter space should be compactified by
identifying $g=+\infty$ with $g=-\infty$, since $H(+ \infty)
=H(-\infty)$. The ground states are an MPS with $A_1=-\sigma^z, A_2=
\sigma^-$ and $A_3= g \sigma^+$~\cite{wolf}, thus $d(g,g') = \sqrt
{1+|g g'|}/ \sqrt {(1+|g|)(1+|g'|)}$ if $g$ and $g'$ are in
different phases, and $d(g,g') = (1+\sqrt {|g g'|})/ \sqrt
{(1+|g|)(1+|g'|)}$ if $g$ and $g'$ are in the same phase. There are
two transition points: $g=0$ and $\infty$. All states for positive
$g$ flow to the state with $g=1$ , and all states for negative $g$
flow to the state with $g=-1$.

The reduced density matrix spectra are trivial, since $\rho =1/2\;I$
for both positive and negative $g$. Therefore, no information is
unveiled from the spectra (and so the von Neumann entropy $E(g)$),
as far as the stable fixed points are concerned. Actually,
$d_E(g,g') =1$, and $E(g)=\ln 2$. The fact that both $d$ and $d_E$
remain the same under interchange $g \leftrightarrow -g$ reflects
that the ground states for $\pm g$ are equivalent up to local
unitaries~\cite{wolf}.


\begin{thebibliography}{99}
\bibitem{ratio} For our purpose, we
restrict to discuss a subsystem consisting of $N=L/2$ consecutive
spins, with $L$ even.
\bibitem{explanation} This is valid for various integrable models, because reduced density
matrix spectra exhibit exponential decay~\cite{spectra} and
degeneracies are the same along RG flows. The exponential character
also occurs in the MPS states, so majorization along RG flows makes
sense.
\bibitem{dual} The model is dual to the disorder-line of the XY model,
i.e., a unit circle given by $\lambda^2 + \gamma^2 =1$ in the
$\lambda-\gamma$ plane, with $\lambda = (1-g^2)/(1+g^2)$ and $
\gamma = 2g/(1+g^2)$.
\end{thebibliography}
\end{document}